\documentclass[useAMS]{mn2e}

\usepackage{graphicx}                   
\input{epsf.sty}                        
\input{psfig.sty}
\begin{document}

\title{GCRT J1745-3009: A precessing radio pulsar?}

\author{W. W. Zhu, R. X. Xu \\
~School of Physics, Peking University, Beijing 100871, China}

\maketitle

\begin{abstract}
A unique transient bursting radio source, GCRT J1745-3009, has
been discovered (Hyman et al. 2005a) near the direction of the
Galactic center.
It is still an open question to explain this phenomenon,
although some efforts to understand its nature have been made.
This paper shows that most of the observed features can be
reproduced by our proposed precessing pulsar model.
It is found that the precession angle of the pulsar should be
larger ($\ga 15^{\rm o}$) than that of previously known precessing
pulsars, which have a precession angle $\la 10^{\rm o}$, if the
beam width of the pulsar is larger than $10$ degree.
The pulsar could be a nulling (or even extremely) radio pulsars to
account for the transient nature of the source.
This model can be confirmed if a pulsar is detected at the
position of the source.
The pulsar could hardly be a normal neutron star (but probably a
solid quark star) if the spin period of the pulsar is detected to
be $\ga 10$ ms in the future.
\end{abstract}


\section{Introduction}

A bursting radio source, GCRT J1745-3009, was discovered at 0.33
GHz in a radio monitoring program of the Galactic
center region made on September 30, 2002 (Hyman et al. 2005a).
Five $\sim 10$ min bursts with peak flux of $\sim 1.67$ Jy were
detected at an apparently regular period of $\sim77$ min from
the source.
Activity (only one single $\sim 0.5$ Jy burst) had been detected
again by GMRT in 330 MHz at 2003 September 28 (Hyman et al.
2005b).
The source appears to be transient because it was not active at
the 1998 September 25 and 26 epochs of VLA observation, and had
not been detected in some other epochs of observation in 2002 and
2003.
Observations indicate that (Hyman et al. 2005b) the burst detected
in 2003 is an isolate one although additional undetected bursts
occurred with 77 min period like the 2002 bursts can not be
completely ruled out.
Assuming that the 2003 burst is an isolated one, Hyman et al.
(2005b) estimated crudely that the duty cycle of the transient
behavior is about 10\%.

Given that (1) the source's brightness temperature would exceed
$10^{12}$ kelvin if it is farther than 100 pc; (2) the source's
observational properties are not directly compatible with that of
any known coherent emitters like white dwarfs or pulsars;
Hyman et al. (2005a) concluded that it it is not likely to be a
incoherent emitter but rather might be one of a new class of
coherent emitter.
Kulkarni \& Phinney (2005) argued that the source could be a
nulling radio pulsar, like PSR J1752+2359 which has quasi periodic
nulling behavior (Lewandowski et al. 2004).
It is pointed out (Turolla, Possenti \& Treves 2005) that the
phenomenon is compatible with what is
expected from the interaction of wind and magnetosphere of two
pulsars in a binary system.
This scenario predicts: (1) a pulsar should be detectable at frequency higher
than 1 GHz; (2) the X-ray luminosity from the shock should be $10^{32}$ ergs s$^{-1}$ which is too low to be detectable by contemporary facilities.
The source could be a white dwarf
(Zhang \& Gil 2005), which may actually behave like a pulsar and
create the activity observed.
This scenario predicts that deep IR exposure with large telescope
may lead to the discovery of the counterpart of GCRT J1745-3009.
A conclusive understanding, however, has not been achieved yet,
and could only be accomplished through further observation.

An alternative effort is tried in this paper to explain the
observational features of GCRT 1745-3009. We propose that the
source could simply be a spinning pulsar precessing with a period
of $\sim 77$ min.
The duration and period of the bursts can be explained with a
broad choice of parameters, as long as the precession angle is not
very small ($>15$ degrees).
It is worth noting that the wobble angle of the pulsar could be
typically of tens of degrees (Melatos 2000) if the free precession
period is close to the radiation-driven precession period.
Given that the brightness temperature could be as high as
$10^{28}-10^{30}$ K, a pulsar could reproduce the observed flux
even if it is as far as 10 kpc away.
The transient nature of the source would be understandable if the
pulsar is an extremely nulling radio pulsar
(Backer 1970; Ritchings 1976; Manchester).
Some of the discovered nulling pulsars could have a huge nulling
fraction. PSR 0826-34 is a case in point, whose nulling fraction
is $70\pm35$ percent (Biggs 1992).
PSR B1931+24 switches off for $\sim 90\%$ of time, and it appears
quasi-periodically at $\sim 40$ days (Cordes et al. 2004; O'Brien
2005).
Such a high fraction of nulling might be consistent with the 10\%
duty-cycle estimated by Hyman et al. (2005b).

A similar idea was presented by Heyl \& Hernquist (2002) who
applied a precessing pulsar model to explain the 6 hours periodic
modulation of X-ray flux from 1E 161348-5055, a neutron star
candidate in the center of the supernova remnant RCW 103.

The model is introduced in \S2. Its application to the pulsar is
discussed in \S3. An extensive discussion on the population of
nulling and precessing pulsars as well as a comparison between our
model and other contemporary models are provided in \S4. The
results are summarized in \S5.

\section{The model}

\begin{figure}
\includegraphics[width=3in]{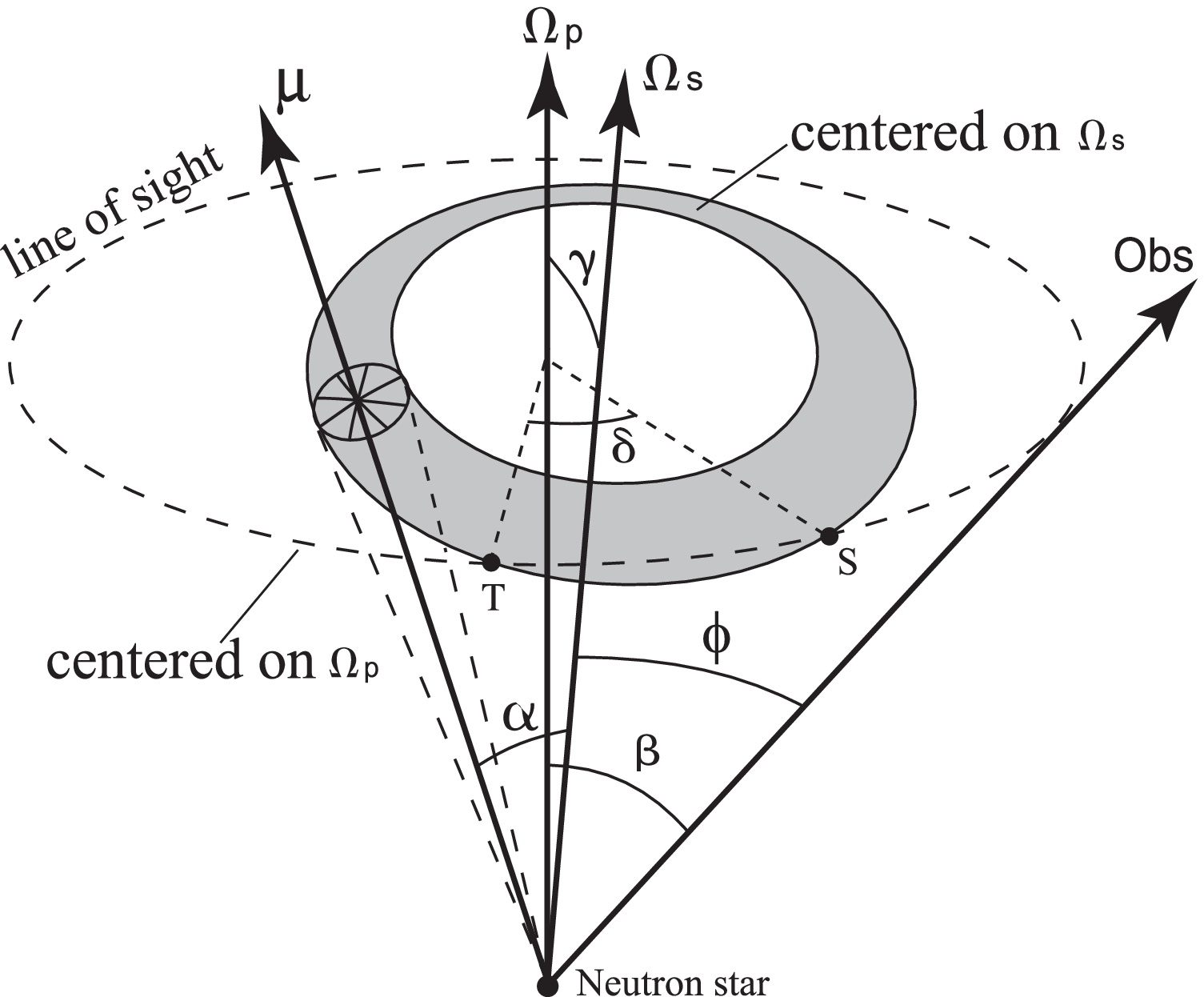}%
\caption{Geometry of a precessing pulsar. $\alpha$ is the angle
between the magnetic axis $\mu$ and the spin axis $\Omega_{\rm
s}$. $\beta$ is the angle between the line of sight ``Obs'' and
the precession axis $\Omega_{\rm p}$. $\phi$ is the angle between
the line of sight and $\Omega_{\rm s}$.
An observer can only detect radio bursts between ``S'' and ``T'',
over an angle $\delta$, which, in our model, is set to be
$\delta=2\pi (10/77)$ to fit the ratio of the observed burst
duration to the period.
\label{fig1}}
\end{figure}

A precessing pulsar scenario is shown in the observer's
rest frame in Fig. \ref{fig1}. The pulsar's
spin axis itself is rotating around a precession axis (which lies
along the direction of total angular momentum).
We denote the magnetic inclination angle as $\alpha$, the angle
between line of sight and the precession axis as $\beta$, and the
precession angle as $\gamma$.
One can also consider another frame, called the precessing
frame, which rotates along $\Omega_{\rm p}$ with the same period
as the precession period.
In this precessing frame, both $\Omega_{\rm p}$ and $\Omega_{\rm
s}$ axes are fixed, and the line of sight rotates about
$\Omega_{\rm p}$.
When the line of sight passes through the emission pattern (shaded
region in Fig. 1), the observer detects burst activity.
The points  ``S'' and ``T''  represent the beginning and end of
the observed burst activity.
$\delta$ is the angle between ``S'' and ``T'' along the trajectory
of the line of sight.

Let's consider the parameter space of $\alpha$, $\beta$, and
$\gamma$, in which the observed flux variation can be successfully
reproduced.
The pulsar's radio emissivity is assumed to be $f(\theta)=f_0
e^{-\theta/\theta_p}$, where $\theta$ is the angular distance from
the magnetic axis $\mu$, and $\theta_p$ is a parameter
characterizing the width of the emission beam.
The observation is sampled every 30 second in the original
observation of Hyman et al. (2005a).
This sampling time is much shorter than the precession period (77 min)
and if it is also longer than the spin period of pulsar,
then the 30s sampled flux, $F_{30}$, can be regarded as a function of $\phi$
(i.e., the angle between line of sight and the pulsar's spin axis).
To simplify the problem, $F_{30}(\phi)$ is assumed to be
proportional to the maximum flux possible in a spin period,
$f(\phi-\alpha)$,\footnote{%
Note that $\phi > \alpha$ if one observes single-peak bursts.
Otherwise, an observer should detect double-peak bursts if $\phi <
\alpha$.
}%
\begin{equation}
F_{30}(\phi_1)/F_{30}(\phi_2)\sim
f(\phi_1-\alpha)/f(\phi_2-\alpha).
\end{equation}
Given that the peak flux observed is $1.67$ Jy, and the undetected
limit is 15 mJy, the ratio of the minimum to maximum fluxes should
thus be $F_{30}(\phi_{\rm max})/F_{30}(\phi_{\rm min})\sim $(15
mJy)/(1.67 Jy)$\simeq 0.01$, where $\phi_{\rm max}$ and $\phi_{\rm
min}$ are the maximum and minimum values of $\phi$ during bursts
(i.e., $F_{30}>15$ mJy), respectively.
Therefore, $f(\phi_{\rm
max}-\alpha)/f(\phi_{\rm min}-\alpha)=\exp[(\phi_{\rm min}-
\phi_{\rm max})/\theta_p]\sim0.01$.
We have then $\phi_{\rm
max}-\phi_{\rm min}=4.7\theta_p$, which is chosen to be $\sim 0.1$
rad $=6^{\rm o}$ since the typical beam width of a normal pulsar
is $\sim 10^{\rm o}$ (Tauris \& Manchester 1998).
The consequence of choosing a larger $\theta_p$ will be discussed
later.
The angle $\delta$ should be set to $\delta=2\pi(10/77)$ in order
to fit the observed ratio of the burst duration to the precession
period.

\begin{figure}
\includegraphics[width=3in]{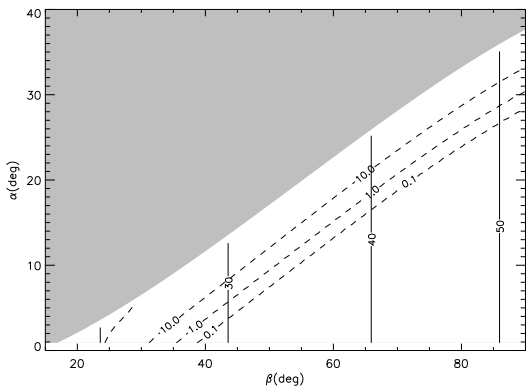}%
\caption{Possible parameter space to reproduce the bursting
behavior in our precession model.
We set here the pulsar's beam radius to be $6^{\rm o}$, the
brightness temperature of radio emission to be $10^{30}$ K, and
the spin angular velocity $\Omega_{\rm s}$ to be 1 rad/s.
The solid and the dashed lines are contours of $\gamma$ (in degrees)
and of source distance (in kpc), respectively,  for given $\alpha$ and $\beta$.
No appropriate $\gamma$ value can be found in the shaded region.
In this calculation, the smallest $\gamma$ we obtain is about $15$
degree in order to reproduce the first five bursts observed during
2002.
\label{fig2}}
\end{figure}

One has $\phi_{\rm min}=\beta-\gamma$ and $\phi_{\rm
max}=\arccos(\cos\beta\cos\gamma+\cos(\delta/2)\sin\beta\sin\gamma)$,
according to spherical geometry. Therefore we have,
\begin{equation}
 \begin{array}{ll}
 \phi_{\rm max}-\phi_{\rm min} &
 =\arccos(\cos\beta\cos\gamma+\cos(\delta/2)
 \sin\beta\sin\gamma) \\
 &+\gamma-\beta  =4.7\theta_p.\\
 \end{array}
\label{thetap}
\end{equation}
The $\gamma$ value can be found from Eq.(\ref{thetap}) for given
$\alpha$, $\beta$, $\theta_p$.
The calculated result is shown in fig \ref{fig2}.
No $\gamma$
solution could be found for $\alpha$ and $\beta$ in the shaded
region in fig 2.
The vertical solid lines in fig \ref{fig2} are the contours of
resulting $\gamma$ from given $\alpha$ and $\beta$ by choosing
$\theta_p=0.1/4.7$ rad.
With the assumption that the pulsar's brightness temperature is
$10^{30}$ K, contours (the dashed lines in Fig. 2) of pulsar
distance can be calculated, provided that the 30s-sampled burst
peak flux is 1.67 Jy.
The distance is computed precisely by simulating the pulsar
emission and integrating the flux over 30 s numerically.
The smallest precessing angle $\gamma$ with which a pulsar can
reproduce the observed bursts is found to be $\sim 16$ degree in
this calculation, while its uncertainty should be $\sim 1$ degree.
Note that the above calculation is based on an assumed structure
of pulsar beam.
Without this assumption, one can also crudely estimate the
smallest possible precessing angle (14 degree if $4.7\theta_p=0.1$
rad and 7 degree for $4.7\theta_p=0.05$ rad) by letting
$\gamma=\beta$ in Eq.(\ref{thetap}).
Thus, we conclude that the pulsar should have a precessing angle
to be larger than $\sim 15$ degree if its beam width is larger
than $10$ degree in our model.

An example of simulated burst profiles is shown in Fig. 3, where
the parameters are $\alpha\simeq 10^{\rm o}$, $\beta\simeq 44^{\rm
o}$, $\gamma\simeq 30^{\rm o}$, and pulsar distance $\simeq 24$
kpc.

\section{The pulsar}

We propose that the enigmatic source, GCRT J1745-3009, could be a
precessing radio pulsar. A radio burst should be detected when the
pulsar's emission beam precesses through the line of sight.
In the model, the distance to the source could be even larger than
10 kpc if the brightness temperature of the pulsar is $\sim
10^{30}$ K.
We find  that the precession angle, $\gamma$, must be rather large
($> \sim 15^{\rm o}$) in order to reproduce the general observed
behavior.
Higher values of the beam radius ($4.7 \theta_p>0.1$) have also been
considered. We find that, as the beam radius increases, the lower
limit of precession angle and the upper limit of the source
distance also increase.

GCRT J1745-3009 was discovered at 0.33 GHz in September 2002, but
was not detected at 1.4 GHz, with a threshold of 35 mJy, in
January 2003 (Hyman et al. 2005a).
Xiang Liu and Huaguang Song also tried to observe the source at 5
GHz with the 25m-radio telescope of the Urumqi station in
Xingjiang, China. They did not detect the source (upper limit of
50 mJy) in observations from 21:20 to 23:55 UT, March 20, 2005,
with an integration time of 30 s.
If the source bursting behavior at 0.33 GHz remained to this
observation, then its spectral index $\alpha$ should be smaller
than $-1.29$.
This value is somewhat smaller than that of the Galactic center
radio transients ($\alpha=-1.2$).
The estimated index ($\alpha < -1.29$) of the source is in
agreement with the typical pulsar spectrum ($\alpha = -1.75$)
obtained using statistics of 285 radio pulsars' spectral indices
between 400 MHz and 1400 MHz (Seiradakis \& Wielebinski 2004).

For a conventional neutron star, a 15-degree precession angle will
induce significantly magnus force and unpin the neutron star's
crust and the superfluid inside (Link \& Cutler 2002).
In this case, the relative deformation of the neutron star crust
is $\epsilon\sim P_s/P_p$, where $P_s$ is the spin period, and
$P_p$ is the 77 min precession period.
Then the deformation will be too large for a conventional neutron
star to have unless the star's spin period is $\sim1$ ms because
Owen (2005) derived the maximum elastic deformation,
$\epsilon_{\rm max}$, of conventional neutron star induced from
shear stresses (Ushomirsky et al. 2000) is only $6.0\times10^{-7}$
(1, fiduciary values of mass, radius are assumed; 2, the breaking
strain is chosen to be $10^{-2}$).
A neutron star with period of 1 ms and $\epsilon\sim 10^{-7}$
would emit gravitational waves that is potentially observable by
long-baseline interferometers like LIGO (Cutler \& Jones 2001;
Melatos \& Payne 2005; Payne \& Melatos 2005 ).
Ostriker \& Gunn (1969) and Heyl \& Hernquist (2002) derived that
the maximum deformation of conventional neutron star induced from
magnetic field is $\epsilon\simeq
4\times10^{-6}(3<B^2_{p,15}>-<B^2_{\phi,15}>)$, where $B_{15}$ is
magnetic field in unit of $10^{15}$G.
This means that if the pulsar have magnetic field like a magnetar
then its period can be as large as tens millisecond.

It is pointed out by Jones \& Andersson (2002) that the upper limit of the
precession angle for a neutron star crust is $\gamma_{\rm max}\sim
0.45(100~{\rm Hz}/f)^2 (u_{\rm break}/10^{-3})$, where $u_{\rm
break}$ is the breaking strain that the solid crust can withstand
prior to fracture .
This means that the value of $u_{\rm break}$ of this pulsar should
be at least $10^{-2}$ which is consistent with the value chosen
for derive the maximum elastic deformation from sheer strain by Owen (2005).
In conclusion, a precessing normal neutron star may reproduce the
observational features only if it is a millisecond pulsar with a
$\sim10^{-2}$ breaking strain.

It has been found that the precession of
normal neutron stars may be damped quickly (in a timescale of
$10^{2\sim 4}$ precession period) via various coupling mechanisms
between the solid crust and the fluid core (Shaham 1977; Levin \& D'Angelo 2004).
If the fast rotating neutron star we are considering here would
also damp that fast ($10^{6-8}$s), then the dissipated energy
(about $\sin\gamma I_{\rm crust}\omega_s^2\sim10^{50}$erg) is too
huge to be unseen in X-ray band (Hyman et al. 2005a).
Therefore, if the bursting activity was produced by a precession
millisecond pulsar, then the pulsar should still be precessing and
possibly detectable by future observation.
If the existence of it is confirmed by future observation then
the damping time scale of large amplitude precessing millisecond
pulsar should be reconsidered.

Alternatively, it is not necessary for the pulsar to rotate very
fast if the pulsar is a solid quark star (Xu 2003, Zhou et al.
2004), because a solid quark star could have a larger elastic
deformation, $\epsilon_{\rm max}\sim 10^{-4}$ (Owen 2005).
Suppose there is no other dissipation mechanism other than
gravitational wave radiation, the typical damping time scale of
precession is $\tau_{\theta}^{\rm rigid}=1.8\times10^6{\rm
yr}(\epsilon/10^{-7})^{-2}(P/0.001\rm s)^4(I/10^{45} \rm g\cdot
cm^2)^{-1} \sim 10^{6-12}\rm yr$ (Bertotti \& Anile 1973; Cutler
\& Jones 2001).
Therefore the bursting activity should remain approximately the same
period and duration provided that it is a solid quark star.

\section{Discussions}

The source had only been observed in activity for twice, the first
time is from 2002 September 30 to October 1, in which five 10-min
duration bursts are detected in a period of 77 minutes (Hyman et
al. 2005a).
The second detection is in 2003 September 28, only one burst is
detected at its decay phase (Hyman et al. 2005b).
The source is likely in quiescent state during other observation
epochs, such as the 1998 September 25 and 26 epochs (Hyman et al.
2005a; 2005b). The sum of the observing time for GCRT J1745-3009
is only 70 hours from 1989 to 2005.
According to this sparse sampling, Hyman et al. (2005b) made their
first crude estimation on the duty-cycle of the source activity
(i.e., $\sim 10\%$).

We propose GCRT J1745-3009 to be a precessing nulling radio pulsar
because of the following reasons.

On the one hand, as we have demonstrated in \S2, a precessing
pulsar with a set of slightly constrained parameters could act
like a bursting radio source if the time resolution of the
observation is not high enough to resolve the pulsar's spin
period. The intriguing source's period and duration, intensity and
distance, as well as the current limitation on its spectra could
be understood in this picture (\S2). The transient nature of the
source could be accounted for if the pulsar is an extremely
nulling pulsar.
Additionally, there is a possible link between the sources and the
supernova remnant since the image of the source shows that the
source is only $10'$ away from the center of a shell-type SNR
G359.1-0.5 (Hyman et al. 2005a; 2005b). The proper motion of the
source further inferred from the supernova's age is $\sim 225 ~\rm
km/s$, namely consistent with the typical kick velocities of
neutron stars.
This later observation supports that GCRT J1745-3009 should be
relevant to neutron stars.

On the other hand, the possibility of existing such a pulsar would
not be too low.
Precessing is rare in pulsars since there are only a few pulsars
which show tentative evidence for precession (Lyne et al. 1998;
Cadez et al. 1997; Jones \& Andersson 2001; Heyl \& Hernquist
2002), i.e., Crab pulsar, Vela pulsar, PSR B1642-03, PSR B1828-11,
the remnant of SN 1987A, Her X-1  and 1E 161348-5055.
Extreme nulling phenomenon with $\sim 10\%$ duty cycle is also not
common for known pulsars.
Within the old data (Biggs 1992), we can only find two pulsars
which show extremely nulling phenomena (PSR 0826-34 and PSR
1944+17).
PSR B1931+24 is suggested to be in a nulling state in about 90\%
of time (Cordes et al. 2004).
Ali (2004) discovered extremely nulling phenomena (nulling
fraction $\sim 70\% - 95\%$) from five pulsars (PSR J1502-5653;
PSR J1633-5102, PSR J1853-0505; PSR J1106-5911; PSR J1738-2335)
and 25 more candidates by analyzing Parkes Multibeam survey data.
Accordingly, one could estimate the possibility of precessing (or
extremely nulling) pulsar to be $7/2000\simeq 0.0035$ since the
total number of discovered pulsars is $\sim 2000$.
Therefore, there should be one precessing and extremely nulling
pulsar in every $10^5$ pulsars if the two phenomena are completely
independent ($0.0035^2\sim 10^{-5}$).
However, the above possibility might have been under-estimated,
because of the following two arguments.
(1). Precessing pulsars and nulling pulsars are more difficult to
detect than ordinary pulsars. Long-term and precise timing is
necessary to confirm precessing phenomena, and special searching
method should be applied to discover an extremely nulling pulsar.
This selection effect should thus reduce significantly the
percentage of these kind of pulsars.
(2). The source's $\sim 10\%$ duty-cycle is a rough estimation
because of the sparse sampling (Hyman et al. 2005b), based on the
assumption that the burst in the 2003 September 28 observation is
isolated.
But, in the second activity, additional undetected bursts other
than the one detected still can't completely be ruled out.
Therefore it is possible that the nulling fraction could be
smaller (even much smaller) than $\sim 90\%$.
Therefore, it could be reasonable for us to detect a radio pulsar
with both precessing and extremely nulling phenomena now.

Is our model less likely than others presented (models of double
neutron star system and of pulsar-like white dwarf)?
Note that the double neutron star model needs also one of the
neutron stars to precess in order to account for the transient
nature. The geodesic precession in the model predicts a 3-year
period of transient behavior, which was not confirmed by the
re-detection of the source in 2003 \cite{hy05b}.
Furthermore, the double neutron star model requires (1) an obital
eccentricity of $\sim0.3-0.6$ in order to change the distance
between the stars significantly; (2) the period of one of the
neutron stars to be close to 0.3 s so that the shock distance from
it can be close to its light-cylinder radius in order to trigger
the ON/OFF switch of the shock emission (Turolla, Possenti \&
Treves 2005).
This would reduce significantly the population of such double
neutron star systems.
Whereas, our model allows the period, the inclination angle (i.e.
$\alpha$) and the angle of line of sight (i.e. $\beta$) to vary in
very large domains.
It could be hasty to conclude that the double neutron star model
is more likely than ours.
The pulsar like white dwarf model presented by Zhang \& Gil (2005)
is interesting.
However, we have never seen any evidence before for the activity
of a pulsar-like white dwarf in the large population of white
dwarfs observed. The peculiarity, origin and population of
pulsar-like white dwarfs need further investigations.

Future observations may uncover the nature of the source.
Predictions for confirming or falsifying our model are provided
below.
It is predicted that a normal or millisecond pulsar should be
detected if the bursting activity is observed in a much higher
timing resolution.

To detect such a pulsar may be a little difficult given the small
duty-cycle of the source and low frequency of the burst activity.
It is said that, in the direction of Galactic center, scattering
would prevent the detection of pulsating radio signal at the
frequency of 330 MHz if the distance of the pulsar is in a range
of ($6\sim 12$) kpc (Turolla, Possenti \& Treves 2005 and
reference therein).
However, it is still possible that pulsing signals could be
observed due to following reasons.
1, The distance of the source could be $<6$ kpc in our model, thus
the scattering effect may be not strong enough to smear the
pulses.
2, It is possible that the pulsar can be detected by some
gamma-ray detector if it has strong magnetospheric activity.
3, Pulsed X-ray emission from the magnetosphere (due to
magnetospheric activity) and/or the surface (due to polar cap
heating) could be high enough to be detected by future instrument
with larger collecting area ($10^{-3}\dot{E}_{\rm rot}$ of a 10 ms
period $10^{12}$ G surface magnetic field pulsar gives $0.2$ mcrab
unabsorbed X-ray flux in a distance of 8 kpc).

In our model, the bursts induced by precession should rise in
almost the same time in different frequencies if the radio beam is
nearly frequency-independent.
One could then observe that the bursting activity begins almost
simultaneously in different channels after DM is considered.
The single pulse searching technique developed by Cordes et al.
(2005) is also a good method to check this prediction.
It is said that this new method is expected to find radio
transients (like GCRT J1745-3009) and a significant number of
pulsars which are not easily identifiable though the period
searching technique (Cordes et al. 2005).

Finally, if the source is a precessing pulsar, its bursts should
be {\em statistically} symmetric since the emissivity of radio
pulsars is generally variable.
If future observation confirm the asymmetric fitting of burst
profile by Hyman et al. 2005a and statistically rule out the
possibility of average symmetric profile, then our model should be
falsified.

If pulsing signals are detected by future observation, one could
distinguish our model from that by Turolla, Possenti \& Treves
(2005) because a precessing pulsar behaves differently from a
pulsar in a binary system in many aspects.
Our model predicts that (1) the frequency shift induced by
precession should be $\Delta\nu/\nu\sim P_s/P_p\sim 10^{-4}$ if
$P_s\sim 0.1$ s (while the shift due to orbital motion in a binary
is $\Delta \nu/\nu\sim 10^{-3}$); (2) pulse width of the pulsar
should vary as the line of sight goes in and out the pulsar's
beam; (3) the timing residual of the pulsar should vary in the
precessing period, with an amplitude of the scale of neutron star
radius\footnote{%
Timing residual and neutron star radius have a same dimension in
case that one set $c=1$.
} %
($<10$ km$/c$, $c$ is the speed of light) which is much smaller
than the timing residual induced by orbital motion ($10^5$
km$/c$).
A fitting to the timing data of observation could distinguish
between these two models.
In summary, it won't be a problem to falsify our model if more
observations are taken in the future.

\section{Summary}

It is shown in this paper that the observed features of GCRT
J1745-3009 can be explained by a precessing nulling radio pulsar
with a precessing angle larger than 15 degrees.
No observation known hitherto could lead one to rule out the model
presented or others (e.g., wind-magnetosphere interaction in
neutron star binary, pulsar like white dwarf).
We also provided some theoretical predictions in the model and
possible ways for falsifying our idea, which could be tested by
future observations.

Discovering of a precessing pulsar with a large precession angle
is interesting, which could provide evidence for a solid quark
star if the pulsar spins at a period of $\ga 10$ ms. This is
certainly very helpful to understand the nature of matter with
supranuclear density.


{\em Acknowledgments}:
The authors thank Dr. Xiang Liu and Mr. Huagang Song for taking
the observation in Nanshan. We are in debt to Prof. Joel Weisberg
for his stimulating comments and suggestions, and for improving
the language. The helpful suggestions from an anonymous referee
are sincerely acknowledged. This work is supported by NSFC
(10273001, 10573002), the Special Funds for Major State Basic
Research Projects of China (G2000077602), and by the Key Grant
Project of Chinese Ministry of Education (305001).


\clearpage

%


\begin{figure}
\includegraphics[width=3in]{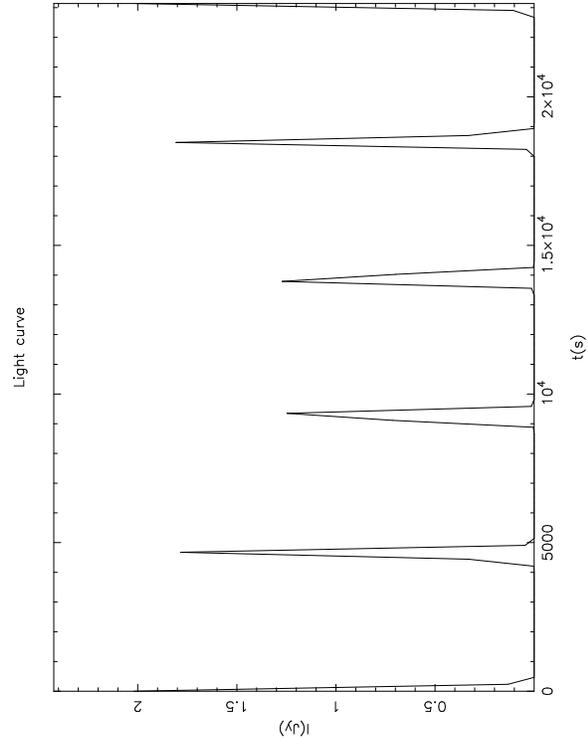}%
\caption{An example of the resulting light curves through
simulation. The parameters chosen are $\alpha\simeq 10^{\rm o}$,
$\beta\simeq 44^{\rm o}$, $\gamma\simeq 30^{\rm o}$, and pulsar
distance $\simeq 24$ kpc. The angular velocity of the spinning
pulsar, $\Omega_{\rm s}$, is set to be 1 rad/s.
\label{fig3}}
\end{figure}

\end{document}